\documentclass[]{sn-jnl}

\usepackage{graphicx}%
\usepackage{multirow}%
\usepackage{amsmath,amssymb,amsfonts}%
\usepackage{amsthm}%
\usepackage{mathrsfs}%
\usepackage[title]{appendix}%
\usepackage{xcolor}%
\usepackage{textcomp}%
\usepackage{manyfoot}%
\usepackage{booktabs}%
\usepackage{algorithm}%
\usepackage{algorithmicx}%
\usepackage{algpseudocode}%
\usepackage{listings}%
\usepackage{graphicx,amsmath,bm}
\usepackage{graphicx}
\usepackage{float}
\usepackage{tablefootnote}
\usepackage{makecell}
\usepackage{float}
\usepackage{subcaption}
\usepackage{pbox}
\usepackage{tikz}
\usetikzlibrary{arrows}
\usepackage{caption}
\usepackage{subcaption}
\usepackage{tikz}
\usetikzlibrary{automata,arrows,positioning,calc}
\usepackage{multicol}
\usepackage{array}
\usetikzlibrary{shapes,fit}
\usepackage{graphicx,amsmath,bm}
\usepackage{graphicx}
\usepackage{float}
\usepackage{tablefootnote}
\usepackage{makecell}
\usepackage{float}
\usepackage{subcaption}
\usepackage{pbox}
\usepackage{caption}
\usepackage{subcaption}
\usepackage{tikz}
\usetikzlibrary{automata,arrows,positioning,calc}
\usepackage{array}
\usetikzlibrary{shapes,fit}
\usepackage[export]{adjustbox}
\theoremstyle{thmstyleone}%
\theoremstyle{thmstyletwo}%

\theoremstyle{thmstylethree}%

\begin{document}

\title[Two Novel Approaches to Detect Community]{Two Novel Approaches to Detect Community: A Case Study of Omicron Lineage Variants PPI Network}
%Innovative Community Detection Methods: Analyzing the PPI Network of Omicron Lineage Variants
%Trailblazing Approaches for Community Detection: A Study of Omicron Lineage Variant's Impact on PPI Network

%%=============================================================%%
%% Prefix	-> \pfx{Dr}
%% GivenName	-> \fnm{Joergen W.}
%% Particle	-> \spfx{van der} -> surname prefix
%% FamilyName	-> \sur{Ploeg}
%% Suffix	-> \sfx{IV}
%% NatureName	-> \tanm{Poet Laureate} -> Title after name
%% Degrees	-> \dgr{MSc, PhD}
%% \author*[1,2]{\pfx{Dr} \fnm{Joergen W.} \spfx{van der} \sur{Ploeg} \sfx{IV} \tanm{Poet Laureate} 
%%                 \dgr{MSc, PhD}}\email{iauthor@gmail.com}
%%=============================================================%%

\author*[]{\fnm{Mamata} \sur{Das}}\email{dasmamata.india@gmail.com}

\author*[]{\fnm{Selvakumar} \sur{K}}\email{kselvakumar@nitt.edu}
%\equalcont{These authors contributed equally to this work.}

\author[]{\fnm{P.J.A.} \sur{Alphonse}}\email{alphonse@nitt.edu}
%\equalcont{These authors contributed equally to this work.}

\affil[]{\orgdiv{Department of Computer Applications}, \orgname{NIT Tiruchirappalli}, \orgaddress{ \city{Trichy}, \postcode{620015}, \state{Tamil Nadu}, \country{India}}}

%\affil[2]{\orgdiv{Department}, \orgname{Organization}, \orgaddress{\street{Street}, \city{City}, \postcode{10587}, \state{State}, \country{Country}}}

%\affil[3]{\orgdiv{Department}, \orgname{Organization}, \orgaddress{\street{Street}, \city{City}, \postcode{610101}, \state{State}, \country{Country}}}

%%==================================%%
%% sample for unstructured abstract %%
%%==================================%%

\abstract{The capacity to identify and analyze protein-protein interactions, along with their internal modular organization, plays a crucial role in comprehending the intricate mechanisms underlying biological processes at the molecular level. We can learn a lot about the structure and dynamics of these interactions by using network analysis. We can improve our understanding of the biological roots of disease pathogenesis by recognizing network communities. This knowledge, in turn, holds significant potential for driving advancements in drug discovery and facilitating personalized medicine approaches for disease treatment. In this study, we aimed to uncover the communities within the variant B.1.1.529 (Omicron virus) using two proposed novel algorithm (ABCDE and ALCDE) and four widely recognized algorithms: Girvan-Newman, Louvain, Leiden, and Label Propagation algorithm. Each of these algorithms has established prominence in the field and offers unique perspectives on identifying communities within complex networks. We also compare the networks by the global properties, statistic summary, subgraph count, graphlet and validate by the modulaity. By employing these approaches, we sought to gain deeper insights into the structural organization and interconnections present within the Omicron virus network.}

\keywords{Omicron, Community Detection, Protein-Protein Interaction}

%%\pacs[JEL Classification]{D8, H51}

%%\pacs[MSC Classification]{35A01, 65L10, 65L12, 65L20, 65L70}

\maketitle

\section{Introduction}\label{sec:introduction}
Community Detection is a process in network analysis where we try to identify groups of nodes that have more connections within the group compared to connections outside the group. It's like finding clusters of friends in a social network where people are more closely connected to others within their cluster than with people outside the cluster. This helps us understand how the network is organized into distinct communities or groups. Across various types of networks, be it social, biological, or dynamic, it is common to observe clusters of nodes that exhibit dense internal connections while displaying sparser connections with the rest of the network. These clusters are known by different names, such as communities \cite{manipur2021community}, modules, clusters \cite{das2021markov}, or even complexes in the context of protein-protein interaction networks (PPINs). Understanding the internal modular organization of PPIs and being able to identify and characterize them through network analysis are pivotal for comprehending molecular-level biological mechanisms \cite{de2010protein}. It is now widely recognized that diseases emerge from disruptions in intricate interactions among extensive sets of genes, rather than being solely attributable to the alteration of a single gene or protein. Consequently, the detection of communities within PPI networks holds promise for unraveling the molecular foundations of disease pathology and facilitating drug discovery efforts. Furthermore, identifying module biological markers specific to individual patients can provide valuable insights for tailoring personalized medical treatments and management approaches.

In the past few years, research efforts concerning community detection (CD) in PPINs derived from disease datasets have concentrated on several key aspects. These include: 
\begin{itemize}
	\item Identifying genes and proteins associated with diseases, with the aim of understanding their involvement in pathological conditions.
	\item  Investigating the tissue specificity of disease modules, shedding light on the specific organs or tissues affected by particular diseases.
	\item Utilizing network-based approaches for disease classification, leveraging the network structure to improve disease diagnosis and categorization.
	\item Predicting potential drug targets and exploring drug repositioning opportunities, utilizing CD methods to identify proteins within disease modules that may serve as effective therapeutic targets.
	\item Identifying biomarkers, which are specific molecular indicators that can be used for disease diagnosis, prognosis, or monitoring treatment response.
\end{itemize}
These research endeavors collectively contribute to our understanding of diseases, facilitate the development of targeted therapies, and pave the way for precision medicine approaches.

\section{Related Work}

Network science has achieved remarkable progress in enhancing our comprehension of complex systems. Within the realm of real-world systems, a key characteristic that emerges is the presence of community structure or clustering. This structure refers to the arrangement of vertices into clusters, where vertices within the same cluster exhibit numerous interconnected edges, while edges connecting vertices from different clusters are relatively sparse. These clusters, often referred to as communities, can be viewed as distinct and relatively autonomous compartments within a graph, akin to the organs or tissues in a human body. Detecting and identifying communities holds immense significance across disciplines such as sociology, biology, and computer science, where systems are commonly represented as graphs. However, this task remains highly challenging and has yet to be fully resolved, despite the concerted efforts of a diverse interdisciplinary community of scientists over the past few years. The complexity and elusive nature of community detection necessitate ongoing research and collaboration to further advance our understanding and develop more robust solutions.

In our study, we have employed hierarchical and semi-supervised clustering methodologies to detect communities within Omicron PPINs. Subsequently, our proposed approach was implemented, comprising two distinct strategies: 1. Average Betweenness-based Community Detection considering Edge (ABCDE) and 2. Average Load-based Community Detection considering Edge (ALCDE). In this section, we will delve into the existing literature regarding community detection in PPINs, focusing on the application of hierarchical approaches such as Girvan-Newman \cite{girvan2002community}, Louvain \cite{blondel2008fast}, and Leiden algorithms \cite{traag2019louvain}, as well as the utilization of semi-supervised clustering methodologies like Label Propagation \cite{raghavan2007near}. By exploring these related works, we aim to provide a comprehensive understanding of the current landscape of community detection in PPINs and highlight the significance of our chosen methodologies.

To detect communities in PPI networks, numerous techniques have been developed. Some of these systems rely purely on network architecture to identify communities, whilst others add biological functions to improve detection. MCODE \cite{bader2003automated}, one of the early algorithms devised for PPI network community detection, provides a localized method by expanding high-ranked nodes (source nodes) into communities. Although MCODE frequently detects enormous communities, it predicts only a small number of true complexes. Another technique, ClusterOne \cite{nepusz2012detecting} detects, employs a greedy strategy that begins with a seed node. It forms communities by iteratively adding or removing nodes with high cohesion. ''ClusterOne" uses an overlapping community discovery method to combine protein groupings that match a specified overlap score threshold. In addition, the Markov Cluster method (MCL) \cite{vandongen2000cluster} is used for PPI network analysis. Using a random walk method, this robust algorithm divides the network into communities.

Recent approaches in PPI network analysis have shown significant improvements by incorporating functional enrichment to accurately detect protein communities. These algorithms capitalize on the observation that protein complexes are typically grouped together to perform specific functions. RNSC \cite{king2004protein} is one of the oldest approaches in this group. It begins with a random partitioning and then optimizes it based on the lowest cost of node exchange. To discover enhanced communities, the algorithm takes into account both density and functional homogeneity, albeit its efficacy can be modified by the initial community assignment. MTGO \cite{vella2018mtgo}, a more modern technique, utilizes both topological and functional information from PPI networks to discover communities. MTGO, like RNSC, starts the process with random partitioning but decides to rejoin nodes into communities if they share common functionality and contribute to enhanced modularity. The method is based on two parameters, min and max, which govern community sizes and influence the final outcomes. Similarly, DCAFP \cite{hu2015density} and GMFTP \cite{zhang2014detecting} are two other algorithms that leverage functional information, but they do not directly involve the biological nature of the networks in their primary process; instead, they preprocess this information through network topology.
% Additionally, we have included a comparison with a flow simulation-based clustering method known as MCL, which was utilized in our previous work \cite{das2023analyzing}.

Hierarchical clustering \cite{johnson1967hierarchical} operates on a bottom-up principle, gradually merging clusters layer by layer. The process commences by considering each individual point as a distinct cluster, subsequently iteratively merging the two clusters with the highest similarity. The measurement of similarity primarily relies on distance metrics, where smaller distances indicate higher similarity. One notable advantage of hierarchical clustering is its ability to discover the hierarchical relationships between classes without requiring the input of a predetermined number of clusters. However, this approach suffers from high time complexity and reduced efficiency. In contrast, k-means \cite{baruri2019comparative} is an early, well-established algorithm widely employed in the field of data mining. Its core strength lies in its significantly lower time complexity compared to hierarchical clustering. By iteratively assigning data points to their nearest cluster centroids, k-means seeks to minimize the within-cluster sum of squares. This iterative process results in the formation of distinct clusters based on the similarity of data points. While k-means offers improved computational efficiency, it does require the user to specify the number of desired clusters in advance.

\paragraph{Girvan-Newman algorithm}The Girvan-Newman algorithm \cite{girvan2002community} is an effective hierarchical approach employed for identifying communities within intricate systems. By systematically eliminating edges from the initial network, the algorithm adeptly uncovers these communities. Rather than solely relying on a metric to determine the central edges within communities, the Girvan-Newman algorithm places emphasis on edges that potentially bridge different communities. It is worth mentioning that the Girvan-Newman algorithm may pose computational challenges when applied to large networks. The calculation of betweenness centrality for all edges demands substantial computational resources, rendering it computationally expensive. Nonetheless, the Girvan-Newman algorithm has found extensive applications across diverse domains, such as social network analysis \cite{latha2023graph}, biological networks \cite{luo2007modular}, and information retrieval \cite{lo2023revisit}. It has proven valuable in detecting communities and investigating the underlying structural characteristics of complex networks. %The Girvan-Newman algorithm for community detection can be outlined as follows:
%\begin{enumerate}
%	\item Calculate the betweenness centrality (BC) for each edge in the graph.
%	\item Remove the edge with the highest BC value.
%	\item Recalculate the BC for the remaining edges.
%	\item Repeat steps 2-3 until no edges remain.
%\end{enumerate}

\paragraph{Louvain Algorithm}The Louvain method \cite{blondel2008fast} is a community detection technique designed to extract communities from extensive networks. This algorithm operates in two phases, namely Modularity Optimization and Community Aggregation. Notably, the Louvain method is an unsupervised approach, meaning it doesn't require prior knowledge of the number or sizes of communities. The algorithm iterates through these two phases repeatedly until no further changes occur in the network and maximum modularity is attained.

\paragraph{Leiden Algorithm}The Leiden algorithm presents an effective approach for detecting communities within vast networks. This algorithm partitions nodes into distinct communities with the primary objective of maximizing the modularity score for each community. Modularity serves as a measure that evaluates the quality of node assignments to communities by comparing the density of connections within a community to that expected in a random network. The Leiden algorithm follows a hierarchical clustering technique, which involves iteratively merging communities into single nodes while optimizing modularity in a greedy manner. This recursive process continues on the condensed graph. A noteworthy aspect of the Leiden algorithm is its enhancement of the Louvain algorithm, addressing specific limitations associated with poorly connected communities \cite{traag2019louvain}. To overcome this issue, the Leiden algorithm incorporates periodic random breakdowns of communities, creating smaller and well-connected sub-communities.

\paragraph{Label Propogation Algorithm}Label propagation is a technique used to assign labels to unlabeled data points in a semi-supervised manner. Initially, a small set of data points with known labels is used. Through iterative steps, labels are spread to unlabeled points based on their similarity or connections. This approach is valuable in revealing community structures within complex networks, where nodes tend to be closely interconnected within specific groups. The algorithm, as outlined in the paper \cite{raghavan2007near} by Raghavan et al., effectively detects these community patterns in large-scale networks. Label propagation presents an algorithmic approach to uncovering communities within such networks. Notably, label propagation offers certain advantages over other algorithms, such as its efficient running time and minimal requirement of prior knowledge regarding the network structure. Unlike some algorithms, it does not rely on known parameters in advance. However, one drawback of label propagation is its lack of producing a unique solution, instead yielding an amalgamation of multiple solutions.

%Label propagation is a semi-supervised machine learning technique that facilitates the assignment of labels to previously unlabeled data points. The algorithm \cite{raghavan2007near} commences with a modest subset of data points possessing labels or classifications. These labels are then propagated to the unlabeled points iteratively during the algorithm's execution. In the realm of complex networks, real-world networks often exhibit a community structure. 
\section{Material and methods}
In this section, our focus is primarily directed towards the dataset, the establishment of PPI networks, and the innovative approach proposed in this research. An overview of the entire workflow is depicted in Fig. \ref{flow}. In our data collection process, we compiled extensive information about the Omicron virus and its lineage variants from the World Health Organization (WHO). Following that, we conducted a search for protein data corresponding to these lineage variants in the UniProtKB database. After obtaining the protein data, we brought together these proteins, combining them through a merging process, and then proceeded to eliminate any duplicate proteins as part of our data cleaning efforts. As a result, we obtained a refined set of distinct proteins, which are potential candidate genes associated with the Omicron virus. The validation of these candidate genes was performed through the STRING database, culminating in the acquisition of protein interaction data. Utilizing this data, we then constructed a PPI network. Subsequently, our proposed approach was implemented, comprising two distinct strategies: 1. Average Betweenness-based Community Detection considering Edge (ABCDE) and 2. Average Load-based Community Detection considering Edge (ALCDE). These novel approaches are comprehensively detailed in subsubection \ref{sec:ABCDE} and subsubection \ref{sec:ALCDE}. To offer a broader perspective, we conducted a comparative analysis by benchmarking our results against well-established algorithms including Girvan Newman, Louvain, Leiden, and Label Propagation. Furthermore, we subjected our network communities to various network comparison constraints, elaborated upon in the Results and Discussion section. The ultimate validation of our community network stemmed from the assessment of its modularity. This intricate process is elaborated upon in the subsequent sections, providing a comprehensive overview of the methodology and the robustness of our findings.

\subsection{Notation and Preliminaries}
%In this study, we consider an undirected and unweighted network denoted as $G = (V, E)$, where $V$ represents the set of nodes, and $E$ represents the set of links. The main objective is to partition $G$ into distinct communities, denoted as $\mathcal{C}$, in such a way that each node $v\in V$ exclusively belongs to one community $C_i$ from the set of communities $\{C_1, C_2, C_3 \cdots, C_n\}$, based on specific connectivity criteria so that $V = \{C_1 \cup C_2 \cup C_3 \cdots \cup C_n\}$
In this study, we take into account an unweighted, undirected network with the notation $\mathcal{G} = (V, E)$, where $V$ stands for the set of nodes and $E$ for the set of links. The main goal is to divide $\mathcal{G}$ into distinct communities, denoted as $\mathcal{C}$, such that each node $v \in V$ exclusively belongs to one community, $C_i$ from the set of communities,  $\{C_1, C_2, C_3 \cdots, C_n\}$, based on certain connectivity criteria, and such that  $V = \{C_1 \cup C_2 \cup C_3 \cdots \cup C_n\}$

\subsection{Dataset}
We utilized an authentic dataset of the Omicron lineage variant, which had been previously employed in our research endeavors \cite{das2023analyzing}. The dataset was obtained from the UniProt/Swiss-Prot database \cite{uniprot2021uniprot}, a reputable source of reviewed and verified information pertaining to proteins within the human body. UniProt/Swiss-Prot serves as a comprehensive protein sequence database, encompassing not only experimental findings and computational features but also scientific conclusions. Its annotations are highly detailed, non-repetitive, and furnish precise functional information about proteins. There is a count of 228 proteins in total initially. We identified and analyzed associated with the following Omicron lineages, see Table \ref{data}. To construct the Omicron Protein-Protein Interaction Network (PPIN), we consolidated the data and meticulously eliminated any duplicate entries. Our curation process involved relying on STRING \cite{snel2000string}, a valuable resource for data validation and the creation of PPINs. The STRING database offers a wealth of information derived from various sources, including computational prediction methods, experimental data, and publicly available text collections. It remains regularly updated and accessible to all free of charge. Furthermore, it employs a spring model to generate network images, treating nodes as masses and edges as springs. Following the data refinement stage, we identified a distinct set of $47$ proteins that constituted the Omicron PPIN. Fig. \ref{47-Omicron} shows the resulting network, showcasing the intricate interactions among the proteins associated with the Omicron lineage.

\subsection{Construction of PPI Network}
\begin{figure}[h]
	\begin{minipage}{0.4\textwidth}\centering
		\includegraphics[width=\textwidth]{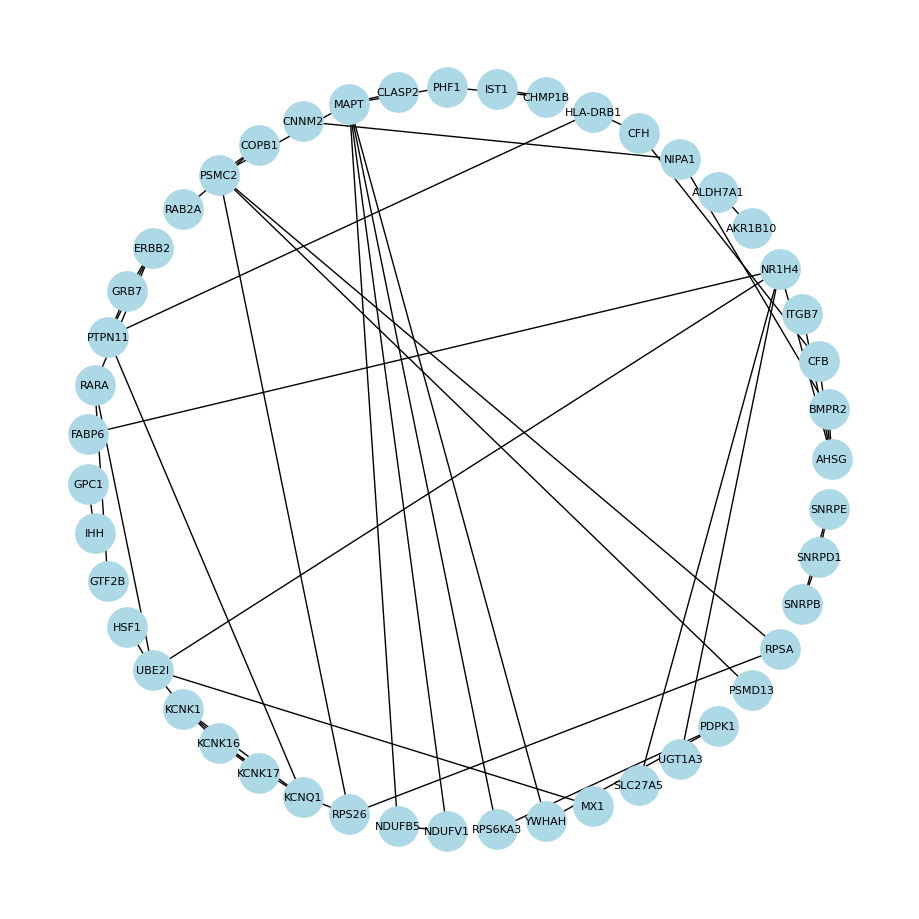}
		\caption{Omicron PPI network}\label{47-Omicron}
	\end{minipage}\hfill
	\begin{minipage}{0.56\textwidth}
		\small
		\captionsetup{singlelinecheck=false}
		\begin{tabular}{@{}lllllll@{}}
			\toprule
			\tiny$B.1.1.529$ & \tiny$BA.5$ & \tiny$BA.4$ & \tiny$BA.3$ & \tiny$BA.2$ & \tiny$BA.1.1$ & \tiny$BA.1$ \\ \midrule
			\tiny27 & \tiny30 & \tiny31 & \tiny34 & \tiny38 & \tiny34 & \tiny34
			\\ \botrule 
		\end{tabular}
		\captionof{table}{Dataset}\label{data}
		%\caption{Dataset}\label{data}
		
		\begin{tabular}{| c | c |}
			\hline
			\#Nodes & 47 \\ 
			\#Edges & 52 \\
			Average degree &  2.213\\
			Minimum degree &  1\\
			Maximum degree &  7\\
			Average Network Clustering &  0.216
			\\ \hline 
			%5 & own &  \\ \hline
		\end{tabular}
		\captionof{table}{Network Structure Properties}
		\label{network-Properties}
		%\caption{Network Structure Properties}
		%\label{network-Properties}
	\end{minipage}
\end{figure}

Our experimental data was gathered from UniProtKB, and we processed the interaction data sourced from STRING. Subsequently, we constructed the Omicron PPI network utilizing NetworkX. Fig. \ref{47-Omicron} presents a circular view of this network, offering a visual representation of its interconnected elements.

\tikzstyle{line} = [draw, -latex']
\begin{figure}[H]
	\tikzstyle{startstop} = [rectangle, rounded corners, minimum width=.6cm, minimum height=.7cm,text centered, draw=black, fill=red!10]
	
	\tikzstyle{process1} = [rectangle, minimum width=11.5cm, minimum height=3cm, text centered, text width=5cm, draw=black, ]
	
	\tikzstyle{process2} = [rectangle, minimum width=3cm, minimum height=2cm, text centered, text width=7.5cm, draw=black, ]
	
	\tikzstyle{process3} = [rectangle, minimum width=2cm, minimum height=2cm, text centered, text width=2.5cm, draw=black, ]
	
	\tikzstyle{process4} = [rectangle, minimum width=2cm, minimum height=2cm, text centered, text width=2.5cm, draw=black,]
	
	\tikzstyle{process5} = [rectangle, minimum width=2cm, minimum height=0.6cm, text centered, text width=4cm, draw=black,]
	
	\tikzstyle{block1} = [rectangle, draw, text width=12em, text centered, rounded corners, minimum height=2em, fill=green!10]
	
	\tikzstyle{block2} = [rectangle, draw, text width=3em, text centered, rounded corners, minimum height=2em, fill=green!10]
	
	\tikzstyle{block3} = [rectangle, draw, text width=5em, text centered, rounded corners, minimum height=2em, fill=green!10]
	
	\tikzstyle{block4} = [rectangle, draw, text width=6em, text centered, rounded corners, minimum height=2em, fill=green!10]
	
	\tikzstyle{block5} = [rectangle, draw, text width=7em, text centered, rounded corners, minimum height=2em, fill=green!10]
	
	\tikzstyle{block6} = [rectangle, draw, text width=6em, text centered, rounded corners, minimum height=2em, fill=green!10]
	
	\tikzstyle{block7} = [rectangle, draw, text width=6em, text centered, rounded corners, minimum height=2em, fill=green!10]
	
	\tikzstyle{block8} = [rectangle, draw, text width=7em, text centered, rounded corners, minimum height=2em, fill=green!10]
	
	\tikzstyle{block9} = [rectangle, draw, text width=6em, text centered, rounded corners, minimum height=2em, fill=green!10]
	
	\tikzstyle{block10} = [rectangle, draw, text width=9em, text centered, rounded corners, minimum height=2em, fill=green!10]
	
	\tikzstyle{block11} = [rectangle, draw, text width=7em, text centered, rounded corners, minimum height=2em, fill=green!10]
	
	\tikzstyle{block12} = [rectangle, draw, text width=6em, text centered, rounded corners, minimum height=2em, fill=green!10]
	
	\tikzstyle{block13} = [rectangle, draw, text width=8em, text centered, rounded corners, minimum height=2em, fill=green!10]
	
	\tikzstyle{block14} = [rectangle, draw, text width=8.5em, text centered, rounded corners, minimum height=2em, fill=green!10]
	
	\tikzstyle{block15} = [rectangle, draw, text width=8em, text centered, rounded corners, minimum height=2em, fill=green!10]
	
	\tikzstyle{block16} = [rectangle, draw, text width=21em, text centered, rounded corners, minimum height=2em, fill=red!10]
	
	\begin{tikzpicture}[node distance = 1.6cm, align=center, auto]
		% Place nodes
		
		\node [startstop] (start) {Start};
		
		\node [process1, below of=start, yshift=-2.5em] (borobox) {Data Collection};
		
		\node [block1, below of=start, ] (a) {Omicron Lineage Varients};
		\node [block2, left of=a, xshift=-5.5em ] (b) {WHO};
		\node [block3, right of=a, xshift=6.5em ] (c) {UniProtKB};
		\node [block4, below of=a,  ] (d) {Data Cleaning};
		\node [block5, below of=b,  ] (e) {Candidate Gene};
		\node [block6, below of=c,  ] (f) {Mearge All};
		
		\node [block7, below of=d,  ] (g) {Collecte PPI Data};
		\node [block8, below of=e,  ] (h) {Validate Protein in STRING};
		\node [block9, below of=f,  ] (i) {Construct PPI Network};
		
		\node [process1, below of=start, yshift=-16.5em] (borobox1) {};
		\node [process2, below of=start, yshift=-17em, xshift=1.5cm,] (borobox2) {Existing Method
			\begin{multicols}{2}
				\begin{itemize}
					\item Grivan Newman
					\item Louvan
				\end{itemize}
				\columnbreak
				\begin{itemize}
					\item Leiden
					\item Label Propagation
				\end{itemize}
		\end{multicols}};
		\node [process3, below of=start, yshift=-17em, xshift=-4cm, ] (borobox3) {Proposed Approach\begin{itemize}
				\item ABCDE
				\item ALCDE
		\end{itemize}};
		
		\node [process1, below of=start, yshift=-30em] (borobox4) {Network Comparison};
		\node [block10, below of=g, yshift=-14em ] (j) {Average Clustering Coefficient};
		\node [block11, left of=j, xshift=-5em] (k) {Network Density};
		\node [block12, right of=j, xshift=5em] (l) {Transitivity};
		\node [block13, below of=j,  ] (m) {Graphlet Analysis};
		\node [block14, left of=m, xshift=-5.6em ] (n) {Degree Distribution};
		\node [block15, right of=m, xshift=5.5em ] (o) {Subgraph Counts};
		
		\node [block16, below of=m,  ] (p) {Validation of Community Networks by Modularity};
		\node [process5, below of=start, yshift=-23.5em] (borobox6) {Communities of Network};
		\path [line] (start) -- (borobox);
		\path [line] (b) -- (a);
		\path [line] (a) -- (c);
		\path [line] (c) -- (f);
		\path [line] (f) -- (d);
		\path [line] (d) -- (e);
		\path [line] (e) -- (h);
		\path [line] (h) -- (g);
		\path [line] (g) -- (i);
		%\path [line] (i) -- (borobox1);
		\path [line] (borobox1) -- (borobox6);
		\path [line] (borobox6) -- (borobox4);
		\path [line] (borobox4) -- (p);
		%\draw [hhhh] (borobox1) -- (d1.center);
		\draw [->] (i.south) -- (borobox1);
	\end{tikzpicture}
	%\caption{Flow chart of the research work}\label{flow}
	\caption{Overview of the process}\label{flow}
\end{figure}
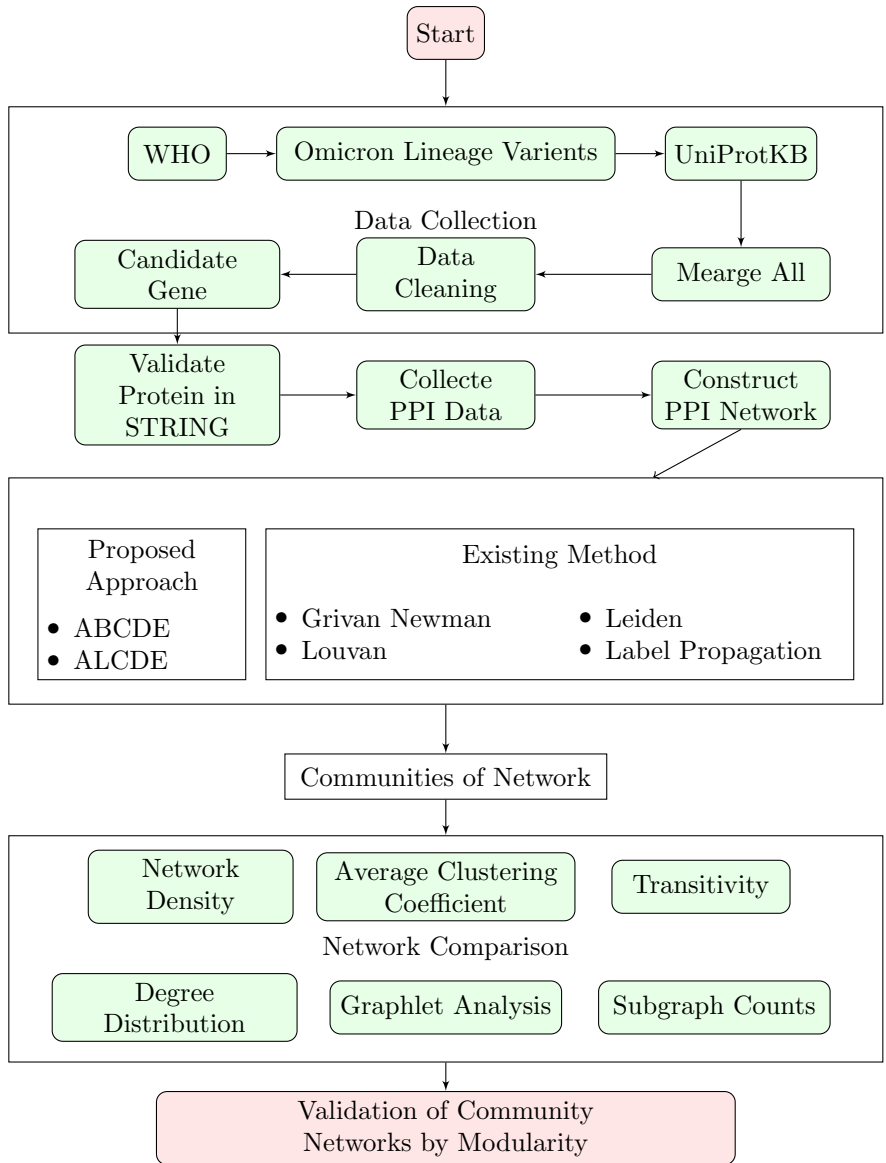

\begin{figure}[H]
	\begin{minipage}[t]{.45\textwidth}
		\centering
		\includegraphics[width=\textwidth]{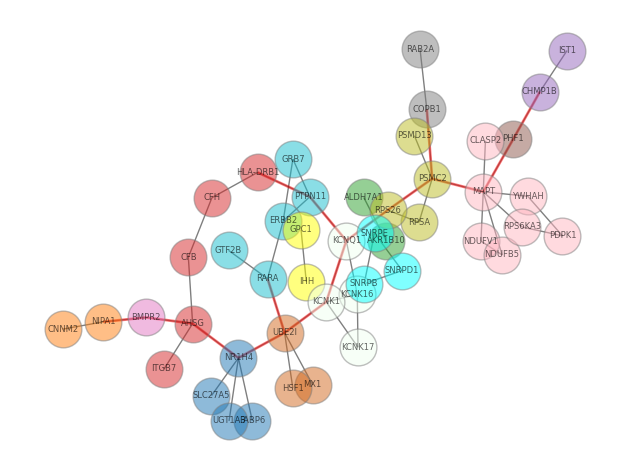}
		\caption{Spring View of Communities in A-EBC}
		\label{edge_bet_c_spring_avg}
	\end{minipage}
\begin{minipage}[t]{.45\textwidth}
	\centering
	\includegraphics[width=\textwidth]{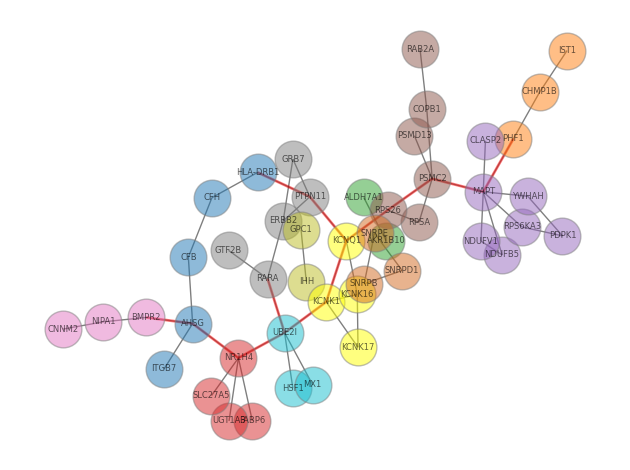}
	\caption{Spring View of Communities in A-ELC}
	\label{edge_load_c_spring_avg}
\end{minipage}
\begin{minipage}[t]{.45\textwidth}
		\centering
		\includegraphics[width=\textwidth]{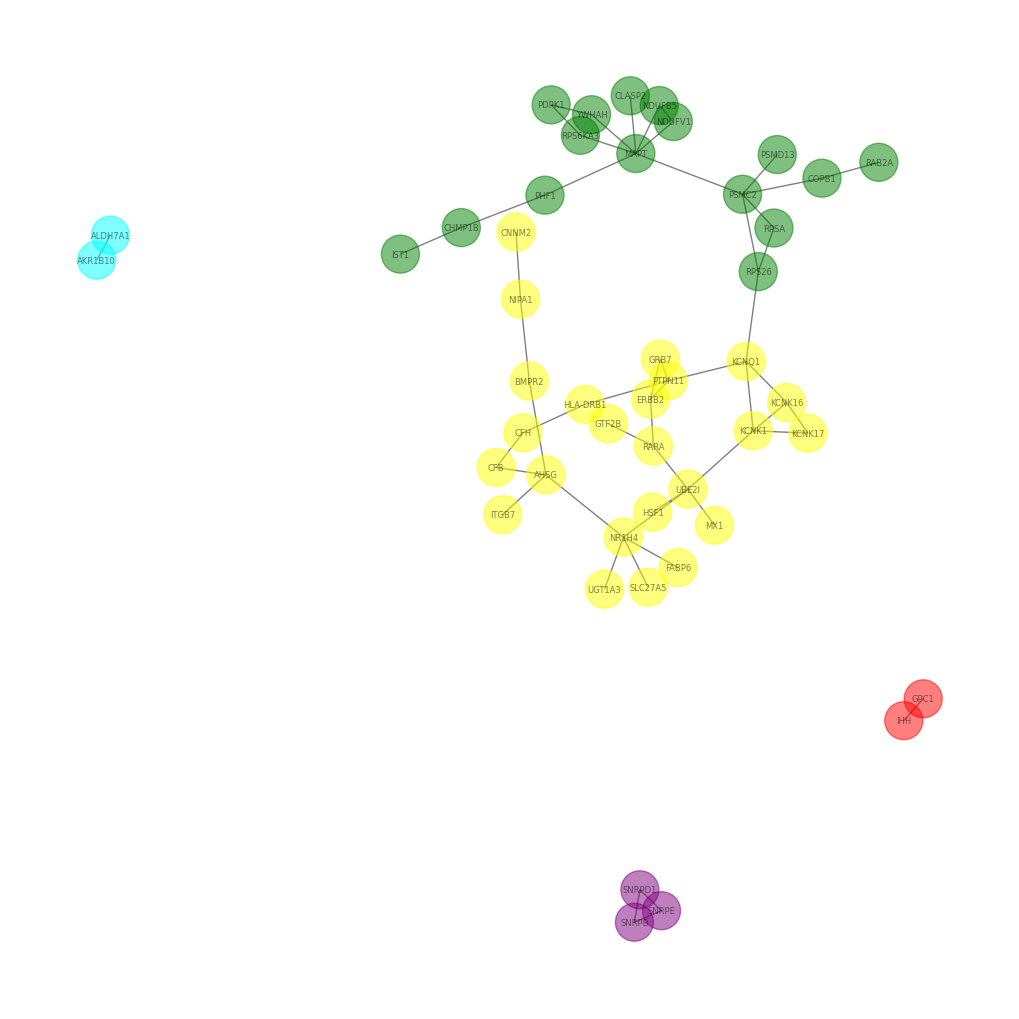}
		\caption{Spring View of Communities in Girvan-Newman Algo}
		\label{griven}
	\end{minipage}
	\begin{minipage}[t]{.45\textwidth}
		\centering
		\includegraphics[width=\textwidth]{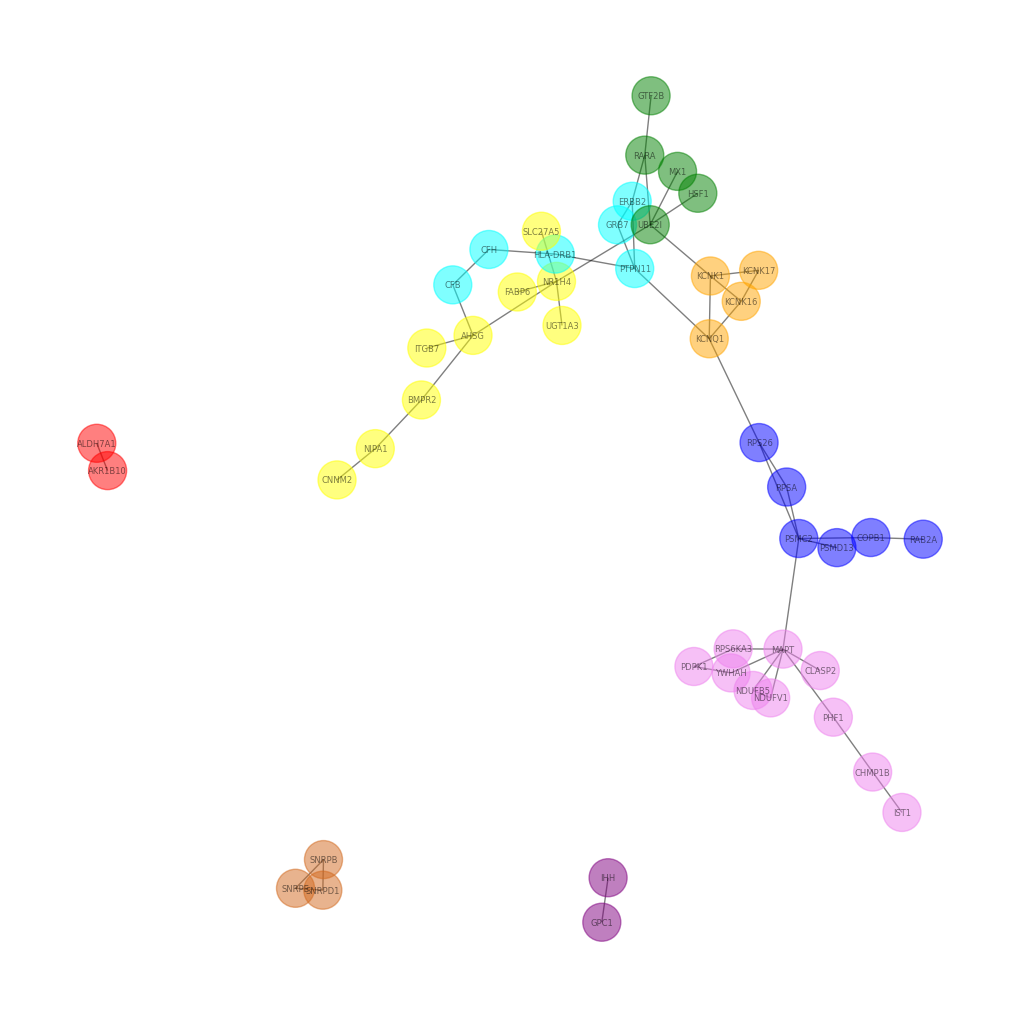}
		\caption{Spring View of Communities in Louvain Algorithm}
		\label{louvain}
	\end{minipage}  
	\begin{minipage}[t]{.45\textwidth}
		\centering
		\includegraphics[width=\textwidth]{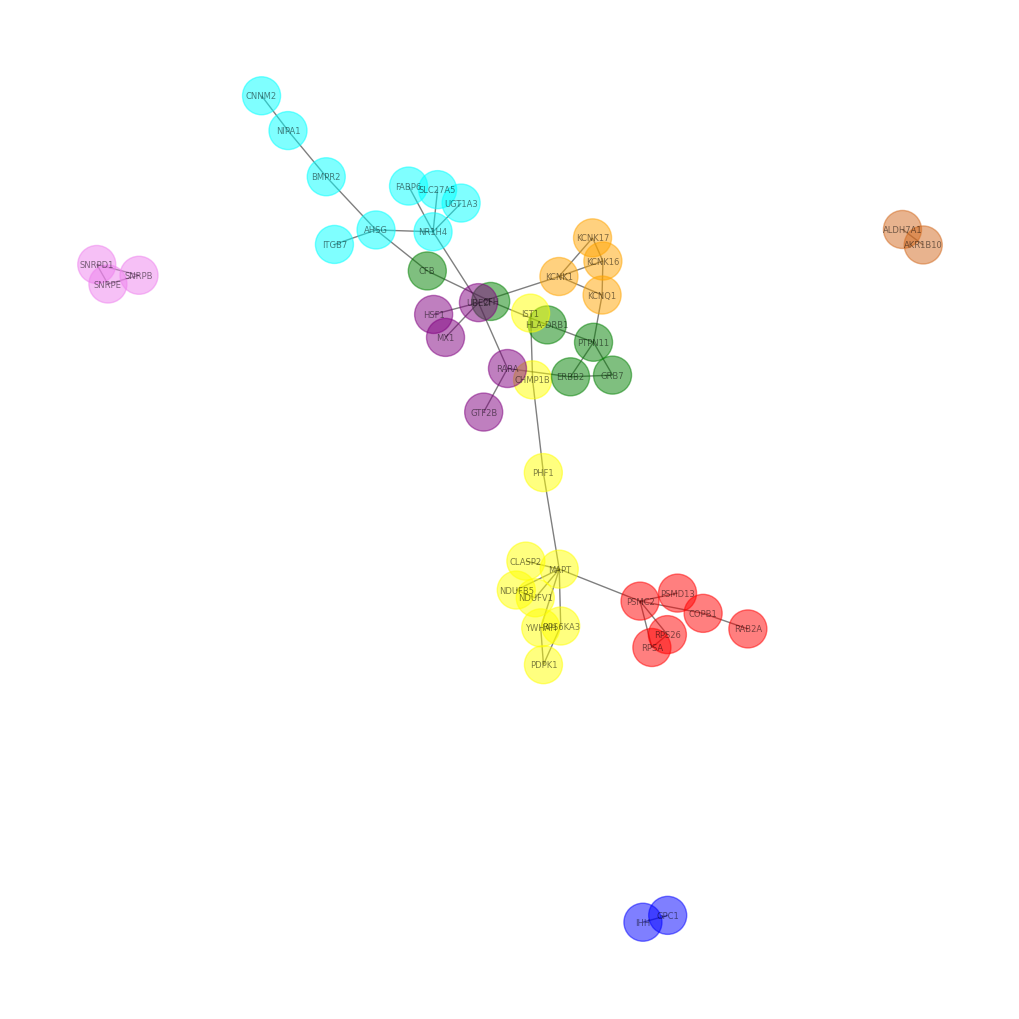}
		\caption{Circular View of Communities in Leiden Algorithm}
		\label{leiden}
	\end{minipage}
	\begin{minipage}[t]{.45\textwidth}
		\centering
		\includegraphics[width=\textwidth]{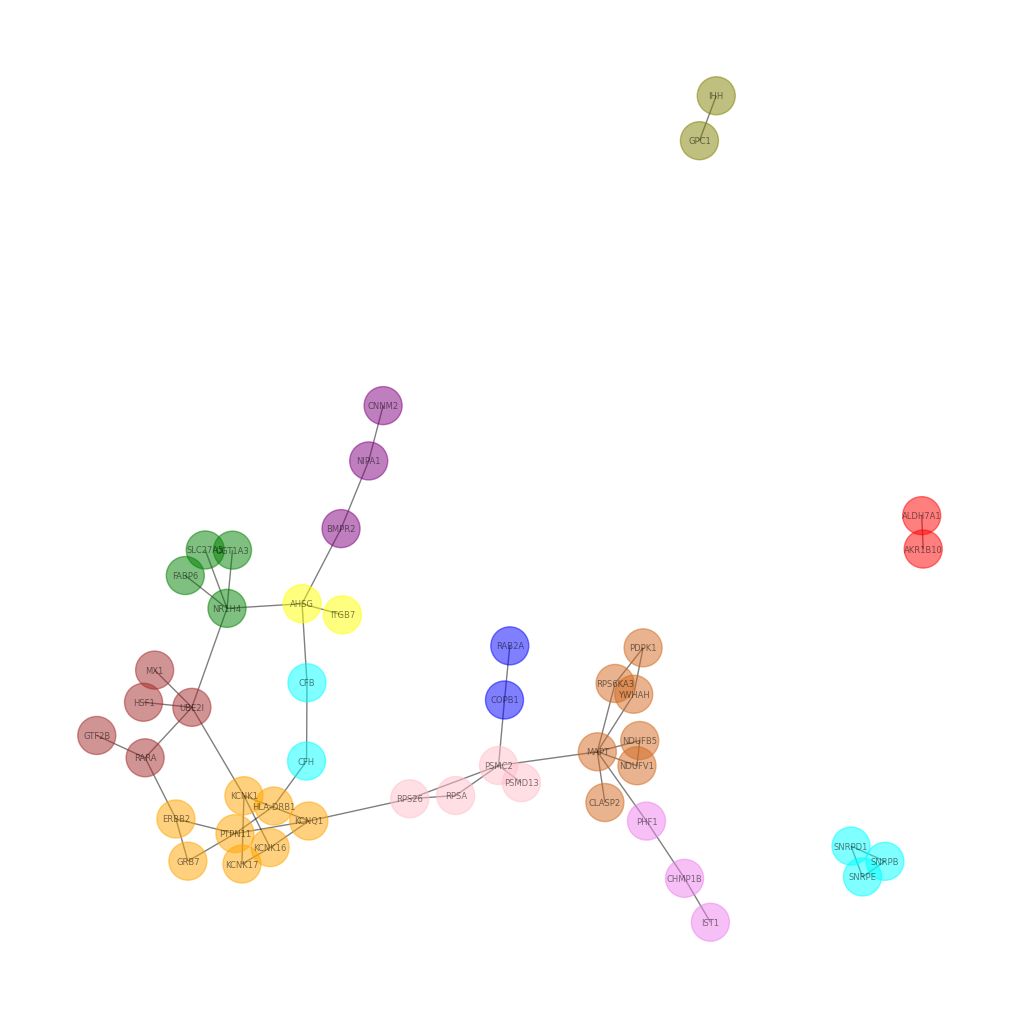}
		\caption{Kamada-kawai View of Communities in Label Propogation Algorithm}
		\label{propo}
	\end{minipage}  
\end{figure}

 Table \ref{network-Properties} illustrates the network structure properties of the Omicron PPI network. To identify communities within the network, we applied four distinct algorithms. The resulting communities were then visualized in Fig. \ref{griven} to Fig. \ref{propo}, using different layout approaches such as circular, Spring, and Kamada-Kawai. We differentiated the communities by assigning different colors to each, facilitating better visualization. To complement the visual representation, we compiled the relevant data into Table \ref{Girvan-Newman_Algorithm_community_detected} to Table \ref{Label-Propogation_Algorithm_community_detected}. These tables provide additional details about the identified communities, aiding in a comprehensive understanding of the relationships and interactions between proteins within the Omicron PPI network. Additionally, in Table \ref{community_detected}, the overall detected communities for each algorithm are displayed.

\subsection{Proposed Approach}

\subsubsection{Algorithm 1: ABCDE}\label{sec:ABCDE}
We have introduced a groundbreaking approach inspired by the concept of edge betweenness \cite{brandes2001faster}. The suggested method, denoted as AEB-CD (Average Edge Betweenness based Community Detection), represents a community detection algorithm. It identifies communities in a network by calculating edge betweenness centrality for each edge. The algorithm iteratively removes edges with edge betweenness greater than or equal to the average threshold value. By removing these critical edges, the algorithm aims to break down the network into smaller components, each representing a community. The process continues until no more edges meet the removal criterion, and the final connected components are the identified communities in the network. Let $\mathcal{G}$ be the graph with vertices $V$ and edges $E$ then our approach can be describe as follows:

\begin{enumerate}
	\item Perform Breadth-First Search (BFS) from node $x$ to find the number of shortest paths from $x$ to each node $v$ in $\mathcal{G}$. Assign the numbers as scores to each node:
	
	Score$(n)$ = number of shortest paths from $x$ to $n$ for all $n \in V$
	
	\item Starting from the leaf nodes, calculate the credit of an edge $(m, n)$ as follows:
	
	\[ Credit(m,n) = ( 1 + \sum_{x({N_m})}^{} Credit(x, m)) \times \frac{\text{Score} (m)}{\text{Score}(n)} \]

	\item Compute the edge credits of all edges in $\mathcal{G}$ using the formula from Step 2. Repeat Step 1 and Step 2 until all nodes are selected.
	
	\item Sum up all the edge credits computed in Step 2 for all edges and divide the total by 2. The result is the edge betweenness of the edges:
	
	\[ E_{B}(m, n)= \frac{1}{2} \sum_{edge(m, n)}^{} Credit(m, n), \forall (m, n)\in E \]
	
	\item Find the average edge betweenness and assign it as the threshold value:
	
	\[Th_{value} = \frac{1}{|E|} \sum_{edge(m, n)}^{} E_{B}(m,n),  \forall (m, n) \in E \]
	
	\item Remove the edges with edge betweenness greater than or equal to the threshold value:
	
	$G^{'}=G$ with edges $(m, n)$ removed if $E_B(m, n) \ge Th_{value}$
	
	\item  Get the communities of the graph $\mathcal{G}^{'}$ after removing the edges. The connected components of $\mathcal{G}^{'}$ represent the communities in the original graph $G$.
\end{enumerate}

\subsubsection{Algorithm 2: ALCDE}\label{sec:ALCDE}
The proposed algorithm ALCDE (Average Load based Community Detection considering Edge), based on average edge load centrality, presents an innovative approach to community detection in networks. By incorporating the concept of random walks and path probabilities, the algorithm efficiently captures the importance of edges in maintaining network connectivity. The use of average edge load centrality as the threshold value for edge removal allows for the identification of well-defined communities within the network. The iterative process of edge removal progressively reveals the underlying community structure, making the algorithm effective and robust in handling diverse network topologies. Overall, the approach shows promise in accurately detecting communities and can be a valuable addition to the study of community detection and network analysis. Given $\mathcal{G}$ is a graph with vertex and edge values of $V$ and $E$.

\begin{enumerate}
	\item Calculate the edge load centrality for each edge $(m, n)$ in $\mathcal{G}$:
	
	\[ E_L(m, n) = \sum_{s\ne m\ne n \ne t}^{} \frac{p_{s \rightarrow t}(m, n)}{p_{s \rightarrow t}}  \]
	where, $p_{s \rightarrow t}(m, n$ )  is the probability that a random walk starting from node s reaches node t by traversing edge $(m, n)$ in one step. $p_{s \rightarrow t}$ is the probability that a random walk starting from node $s$ reaches node $t$.
	
	\item Calculate the average edge load centrality value as the threshold value:
	
	\[ Th_{value} = \frac{1}{|E|} \sum_{(m, n) \in E}^{} E_L(m, n) \]
	where $|E|$ is the total number of edges in $\mathcal{G}$.
	
	\item Remove the edges with edge load centrality greater than or equal to the threshold value:
	
	$G^{'}=G$ with edges $(m,n)$ removed if $E_L(m,n) \ge Th_{value}$
	
	\item  Get the communities of the graph $\mathcal{G}^{'}$ after removing the edges. The connected components of $\mathcal{G}^{'}$ represent the communities in the original graph $\mathcal{G}$.
\end{enumerate}

This algorithm uses edge load centrality to identify the importance of each edge in facilitating communication between nodes in the network. By removing edges with high edge load centrality (greater than or equal to the threshold), the algorithm breaks down the network into smaller connected components, each representing a community. The process continues until no more edges meet the removal criterion, and the final connected components are the identified communities in the network.

\section{Results and Discussions}\label{sec:results}

\begin{table}[h]
	\caption{Edge betweenness value}
	\label{edge_betweenness}
	\centering
	\begin{tabular}{@{}llll@{}}
		\toprule
		\tiny Edges & \tiny Edge Betweenness
		& \tiny Edges &\tiny Edge Betweenness
		\\ \midrule
		 %KCNQ1 -- RPS26 &  0.355226642 & 	 NIPA1 -- CNNM2 & 0.036077706 \\
		\tiny KCNQ1 -- RPS26 & \tiny 0.355226642 & 	\tiny NIPA1 -- CNNM2 & \tiny 0.036077706 \\
		\tiny PSMC2 -- RPS26 & \tiny 0.323774283 & \tiny CHMP1B -- IST1 & \tiny 0.036077706 \\
		\tiny MAPT -- PSMC2 & \tiny 0.277520814 & \tiny CLASP2 -- MAPT & \tiny 0.036077706 \\
		\tiny UBE2I -- KCNK1 & \tiny 0.251002158 & \tiny COPB1 -- RAB2A & \tiny 0.036077706 \\
		\tiny KCNK1 -- KCNQ1 & \tiny 0.238976257 & \tiny PSMC2 -- PSMD13 & \tiny 0.036077706 \\
		\tiny NR1H4 -- UBE2I & \tiny 0.235892692 & \tiny RARA -- GTF2B & \tiny 0.036077706\\
		\tiny AHSG -- NR1H4 & \tiny 0.161887142 & \tiny HSF1 -- UBE2I & \tiny 0.036077706\\
		\tiny PTPN11 -- KCNQ1 & \tiny 0.128276287 & \tiny UBE2I -- MX1 & \tiny 0.036077706\\
		\tiny AHSG -- BMPR2 & \tiny 0.102682701 & \tiny MAPT -- NDUFB5 & \tiny 0.035152636\\
		\tiny PHF1 -- MAPT & \tiny 0.102682701 &  \tiny MAPT -- NDUFV1 & \tiny 0.035152636\\
		\tiny HLA-DRB1 -- PTPN11 & \tiny 0.07493062 &  \tiny KCNK16 -- KCNQ1 & \tiny 0.030835646\\
		\tiny RARA -- UBE2I & \tiny 0.070613629 & \tiny KCNK1 -- KCNK17 & \tiny 0.025131051\\
		\tiny BMPR2 -- NIPA1 & \tiny 0.070305273 & \tiny GRB7 -- PTPN11 & \tiny 0.024514339\\
		\tiny CHMP1B -- PHF1 & \tiny 0.070305273 & \tiny RPS26 -- RPSA & \tiny 0.023126735\\
		\tiny COPB1 -- PSMC2 & \tiny 0.070305273 & \tiny RPS6KA3 -- PDPK1 & \tiny 0.018501388\\
		\tiny CFH -- HLA-DRB1 & \tiny 0.05920444 & \tiny YWHAH -- PDPK1 & \tiny 0.018501388\\
		\tiny MAPT -- RPS6KA3 & \tiny 0.051803885 & \tiny KCNK1 -- KCNK16 & \tiny 0.014338575\\
		\tiny MAPT -- YWHAH & \tiny 0.051803885 &  \tiny PSMC2 -- RPSA & \tiny 0.012950971\\
		\tiny ERBB2 -- RARA & \tiny 0.050262103 & \tiny ERBB2 -- GRB7 & \tiny 0.011563367\\
		\tiny ERBB2 -- PTPN11 & \tiny 0.047332717 & \tiny KCNK16 -- KCNK17 & \tiny 0.010946654\\
		\tiny CFB -- CFH & \tiny 0.047178538 & \tiny AKR1B10 -- ALDH7A1 & \tiny 0.000925069\\
		\tiny AHSG -- CFB & \tiny 0.041628122 & \tiny GPC1 -- IHH & \tiny 0.000925069\\
		\tiny AHSG -- ITGB7 & \tiny 0.036077706 & \tiny NDUFB5 -- NDUFV1 & \tiny 0.000925069\\
		\tiny NR1H4 -- FABP6 & \tiny 0.036077706 & \tiny SNRPB -- SNRPD1 & \tiny 0.000925069\\
		\tiny NR1H4 -- SLC27A5 & \tiny 0.036077706 & \tiny SNRPB -- SNRPE & \tiny 0.000925069\\
		\tiny NR1H4 -- UGT1A3 & \tiny 0.036077706 & \tiny SNRPD1 -- SNRPE & \tiny 0.000925069\\
		%row 3    & data 7   & data 8  & data 9\footnotemark[2]  \\
		\botrule
	\end{tabular}
\end{table}
%\footnotetext{Source: This is an example of table footnote. This is an example of table footnote.}
%\footnotetext[1]{Example for a first table footnote. This is an example of table footnote.}
%\footnotetext[2]{Example for a second table footnote. This is an example of table footnote.}

\begin{table}
	\parbox{.45\linewidth}{
		\centering
		\caption{Comparision of network's global properties in different methods} \label{compare}
		\begin{tabular}{@{}llll@{}}
			\toprule
			Method &  $\mathcal{D}$ &  $\mathcal{T}$ & $\mathcal{C}_{coef}$\\ 
			\midrule
			Omicron PPIN & 0.048 & 0.170 & 0.216 \\
			A-EBL & 0.034 & 0.315 & 0.208 \\
			A-ELC &  0.037 & 0.279 & 0.208\\
			Grivan Newman & 0.047 & 0.178 & 0.234 \\
			Louvain & 0.042 & 0.281 & 0.274 \\
			Leiden & 0.042 & 0.281 & 0.274 \\
			Label Propagation &  0.038 & 0.36  & 0.261\\
			\botrule
		\end{tabular}
	}
	\hfill
	\parbox{.45\linewidth}{
		\centering
		\caption{No. of Communities detected}
		\begin{tabular}{@{}lll@{}}
			\toprule
			Sl. No. & Algorithms & \makecell{\#Communities} \\ \midrule
			1 & AT-EBC & 14 \\
			2 & AT-ELC & 12 \\
			3 & Girvan-Newman &  5 \\ %\hline
			4 & Louvain & 9\\ %\hline
			5 & Leiden & 9 \\ %\hline
			6 & Label Propogation & 13 \\ 
			%7 & MCL & 18\\
			\botrule 
			%5 & own &  \\ \hline
		\end{tabular}
	}
\end{table}

\begin{table}[h]
	\caption{No. of Communities detected in A-EBC}
	\label{A-EBC_community_detected}
	\centering
	\begin{tabular}{@{}lll@{}}
		\toprule
		$\mathcal{C}$ &  $\mathcal{P}_{count}$ &  $\mathcal{P}_{names}$ \\ 
		\midrule
		$\mathcal{C}_1$ & 5 & \tiny{'ITGB7', 'CFB', 'HLA-DRB1', 'AHSG', 'CFH'}\\
		$\mathcal{C}_2$ & 4 & \tiny{'NR1H4', 'FABP6', 'SLC27A5', 'UGT1A3'}\\
		$\mathcal{C}_3$ & 2 & \tiny{'AKR1B10', 'ALDH7A1'}\\
		$\mathcal{C}_4$ & 1 & \tiny{'BMPR2'}\\
		$\mathcal{C}_5$ & 2 & \tiny{'CNNM2', 'NIPA1'}\\
		$\mathcal{C}_6$ & 2 & \tiny{'CHMP1B', 'IST1'}\\
		$\mathcal{C}_7$ & 1 & \tiny{'PHF1'}\\
		$\mathcal{C}_8$ & 7 & \tiny{'MAPT', 'NDUFV1', 'YWHAH', 'CLASP2', 'PDPK1', 'RPS6KA3', 'NDUFB5'}\\
		$\mathcal{C}_9$ & 2 & \tiny{'RAB2A', 'COPB1'}\\
		$\mathcal{C}_{10}$ & 4 & \tiny{'PSMD13', 'RPSA', 'RPS26', 'PSMC2'}\\
		$\mathcal{C}_{11}$ & 5 & \tiny{'GTF2B', 'GRB7', 'ERBB2', 'RARA', 'PTPN11'}\\
		$\mathcal{C}_{12}$ & 2 & \tiny{'IHH', 'GPC1'}\\
		$\mathcal{C}_{13}$ &3 & \tiny{'HSF1', 'MX1', 'UBE2I'}\\
		$\mathcal{C}_{14}$ & 4& \tiny{'KCNQ1', 'KCNK16', 'KCNK1', 'KCNK17'}\\
		$\mathcal{C}_{15}$ & 3 & 	\tiny{'SNRPB', 'SNRPE', 'SNRPD1'}\\
		\botrule
	\end{tabular}
\end{table}

\begin{table}[h]
	\caption{No. of Communities detected in AT-ELC}
	\label{A-ELC_community_detected}
	\centering
	\begin{tabular}{@{}lll@{}}
		\toprule
		$\mathcal{C}$ &  $\mathcal{P}_{count}$ &  $\mathcal{P}_{names}$ \\ 
		\midrule
		$\mathcal{C}_1$ & 5 & \tiny{HLA-DRB1, CFH, CFB, AHSG, ITGB7}\\
		$\mathcal{C}_2$ & 4 & \tiny{SLC27A5, NR1H4, FABP6, UGT1A3}\\
		$\mathcal{C}_3$ & 2 & \tiny{ALDH7A1, AKR1B10}\\
		$\mathcal{C}_4$ & 3 & \tiny{BMPR2, CNNM2, NIPA1}\\
		$\mathcal{C}_5$ & 3 & \tiny{IST1, CHMP1B, PHF1}\\
		$\mathcal{C}_6$ & 7 & \tiny{RPS6KA3, CLASP2, NDUFV1, MAPT, PDPK1, YWHAH, NDUFB5}\\
		$\mathcal{C}_7$ & 6 & \tiny{PSMD13, RPS26, RPSA, RAB2A, PSMC2, COPB1}\\
		$\mathcal{C}_8$ & 5 & \tiny{PTPN11, GRB7, RARA, ERBB2, GTF2B}\\
		$\mathcal{C}_9$ & 2 & \tiny{IHH, GPC1}\\
		$\mathcal{C}_{10}$ & 3 & \tiny{HSF1, MX1, UBE2I}\\
		$\mathcal{C}_{11}$ & 4 & \tiny{KCNK16, KCNK17, KCNK1, KCNQ1}\\
		$\mathcal{C}_{12}$ & 3 & \tiny{SNRPD1, SNRPB, SNRPE}\\
		 \botrule
	\end{tabular}
\end{table}

\begin{table}[h]
	\caption{No. of Communities detected in Girvan-Newman}
	\label{Girvan-Newman_Algorithm_community_detected}
	\centering
	\begin{tabular}{@{}lll@{}}
		\toprule
		$\mathcal{C}$ &  $\mathcal{P}_{count}$ &  $\mathcal{P}_{names}$ \\ \midrule
		1 & 24 &\tiny\makecell{CFH, HLA-DRB1, GRB7, UGT1A3, CNNM2, MX1, PTPN11, AHSG, NIPA1, KCNK1,\\ HSF1, ITGB7, FABP6, RARA, CFB, KCNK16, BMPR2, GTF2B, UBE2I, KCNK17,\\ SLC27A5, NR1H4, ERBB2, KCNQ1}  \\ %\hline
		2 & 2 & \tiny{AKR1B10, ALDH7A1} \\ %\hline
		3 & 16 & \tiny\makecell{NDUFB5, PSMD13, NDUFV1, PSMC2, PDPK1, RAB2A, COPB1, RPS6KA3, PHF1, \\ CHMP1B, YWHAH, RPS26, CLASP2, IST1, MAPT, RPSA}\\ %\hline
		4 &  2 & \tiny{GPC1, IHH}\\ %\hline 
		5 & 3  & \tiny{SNRPB, SNRPD1, SNRPE}\\ 
		\botrule
	\end{tabular}
\end{table}

\begin{table}[h]
	\caption{No. of Communities detected in Louvain Algorithm}
	\label{Louvain_Algorithm_community_detected}
	\centering
	\begin{tabular}{@{}lll@{}}
		\toprule
		$\mathcal{C}$ &  $\mathcal{P}_{count}$ &  $\mathcal{P}_{names}$ \\ 
		\midrule
		1 & 6 & \tiny{CFB ,CFH ,HLA-DRB1 ,ERBB2 ,GRB7 ,PTPN11} \\
		2 & 5 & \tiny{RARA ,GTF2B ,HSF1 ,UBE2I ,MX1} \\
		3 & 2 & \tiny{GPC1 ,IHH} \\
		4 & 2 & \tiny{AKR1B10 ,ALDH7A1}\\
		5 & 9 & \tiny{AHSG ,ITGB7 ,NR1H4 ,BMPR2 ,NIPA1 ,CNNM2 ,FABP6 ,SLC27A5 ,UGT1A3}\\
		6 & 4 & \tiny{KCNK1 ,KCNK16 ,KCNK17 ,KCNQ1}\\
		7 & 10 & \tiny{CHMP1B ,IST1 ,PHF1 ,CLASP2 ,MAPT ,NDUFB5 ,NDUFV1 ,RPS6KA3 ,YWHAH ,PDPK1}\\
		8 & 3 & \tiny{SNRPB ,SNRPD1 ,SNRPE}\\
		9 & 6 & \tiny{COPB1 ,PSMC2 ,RAB2A ,RPS26 ,PSMD13 ,RPSA} \\ \botrule
	\end{tabular}
	%\newline \footnotesize{3 Communities detected by Louvain Algorithm}\\
	%\newline \footnotesize{$^1$3 Communities detected by ...}\\
\end{table}

\begin{table}[h]
	\caption{No. of Communities detected in Leiden Algorithm}
	\label{Leiden_Algorithm_community_detected}
	\centering
	\begin{tabular}{@{}lll@{}}
		\toprule
		$\mathcal{C}$ &  $\mathcal{P}_{count}$ &  $\mathcal{P}_{names}$ \\ 
		\midrule
		1  & 10 & \tiny \pbox{30cm}{CHMP1B, IST1, PHF1, CLASP2, MAPT, NDUFB5, NDUFV1, RPS6KA3, YWHAH, PDPK1} \\ 
		2  & 9 & \tiny{AHSG, ITGB7, NR1H4, BMPR2, NIPA1, CNNM2, FABP6, SLC27A5, UGT1A3} \\ 
		3  & 6 & \tiny{COPB1, PSMC2, RAB2A, RPS26, PSMD13, RPSA} \\ %\hline
		4  & 6 & \tiny{CFB, CFH, HLA-DRB1, ERBB2, GRB7, PTPN11} \\ %\hline
		5  & 5 & \tiny{RARA, GTF2B, HSF1, UBE2I, MX1}\\
		6  & 4 & \tiny{KCNK1, KCNK16, KCNK17, KCNQ1} \\ %\hline
		7  & 3 & \tiny{SNRPB, SNRPD1, SNRPE} \\ %\hline
		8  & 2 & \tiny{AKR1B10, ALDH7A1} \\ %\hline
		9  & 2  & \tiny{GPC1, IHH} \\ 
		\botrule
	\end{tabular}
\end{table}

\begin{table}[h]
	\caption{No. of Communities detected in Label Propogation Algorithm}
	\label{Label-Propogation_Algorithm_community_detected}
	\centering
	\begin{tabular}{@{}lll@{}}
		\toprule
		$\mathcal{C}$ &  $\mathcal{P}_{count}$ &  $\mathcal{P}_{names}$  \\ 
		\midrule
		1 & 2 & \tiny{AHSG, ITGB7}  \\% \hline
		2 & 3 & \tiny{NIPA1, BMPR2, CNNM2}  \\ %\hline
		3 & 2 & \tiny{CFH, CFB}  \\ %\hline
		4 & 4 & \tiny{NR1H4, FABP6, UGT1A3, SLC27A5}  \\ %\hline
		5 & 2 & \tiny{AKR1B10, ALDH7A1}  \\% \hline
		6 & 8 & \tiny{HLA-DRB1, KCNK1, GRB7, KCNK17, PTPN11, ERBB2, KCNK16, KCNQ1}  \\ %\hline
		7 & 3 & \tiny{PHF1, CHMP1B, IST1 } \\ %\hline
		8 & 7 & \tiny{NDUFB5, NDUFV1, PDPK1, RPS6KA3, YWHAH, CLASP2, MAPT}  \\ %\hline
		9 & 2 & \tiny{RAB2A, COPB1}  \\ %\hline
		10 & 4 & \tiny{PSMD13, RPSA, PSMC2, RPS26}  \\ %\hline
		11 & 5 & \tiny{GTF2B, UBE2I, HSF1, RARA, MX1}  \\ %\hline
		12 & 2 & \tiny{GPC1, IHH}  \\ %\hline
		13 & 3 & \tiny{SNRPB, SNRPD1, SNRPE}  \\ 
		\botrule
	\end{tabular}
\end{table}

\subsection{Network Comparison}
Networks can be compared using various measures like network density, degree distribution, transitivity, and clustering coefficient, which provide an overview of their global properties. Another approach to compare networks is by analyzing subgraph \cite{rito2010threshold} and graphlets \cite{prvzulj2007biological} counts. These measures involve counting small subgraphs, such as triangles, stars, squares, and cliques, within the networks. By comparing the frequencies of these subgraphs, we can gauge the similarity between networks based on their subgraph structures. To compare networks, we create frequency vectors of subgraphs and then assess the similarities in subgraph counts between the networks. This gives us an idea of how the networks relate to each other in terms of their subgraph patterns. Additionally, we have compare networks through the concept of modularity of community to validate the networks. %This approach is often used to evaluate community structures in networks generated using different methods.
Despite the availability of various network comparison techniques, declaring two networks as similar, different, or quantifying their correlation is not straightforward. Each method comes with its own unique challenges, such as dealing with networks originating from the same model but having different node properties. Thus, network comparison remains a complex task with diverse aspects to consider.

\subsubsection{Network Density}
Network density\cite{army2006field} is a metric used to measure how closely connected the nodes are in a network. When comparing community networks, we use network density to understand the level of interconnections within each community. A higher network density indicates stronger connections, suggesting more tightly-knit and cohesive communities. On the other hand, lower network density may indicate more loosely connected communities with fewer interactions among their nodes. By analyzing network density, we gain insights into the structural variations and organization of communities across different networks. Network density (D) for a community in a graph can be mathematically defined as:

\begin{equation}
	\mathcal{D} = \frac{2b}{a(a-1)}
	\label{network_density}
\end{equation}
where the number of nodes is $a$ and the number of edges is $b$. For a complete graph or network, density is $1$, and $0$ for a graph without any edges. We have refered 2nd column of Table \ref{compare} to observe the density values for various community networks generated by different methods. %These density values provide valuable insights into the structural characteristics of each community network, aiding in a better understanding of their network.

\subsubsection{Transitivity}
Transitivity\cite{holland1971transitivity} is important for comparing community networks because it shows how closely connected the nodes are within each group. Higher transitivity means the nodes in a community are tightly linked, making it a stronger and more connected group. Comparing transitivity values helps us find similar patterns and overlapping groups in different networks. Mathematically, 

\begin{equation}
	\mathcal{T} = 3 \times \frac{\#c}{\#d}
	\label{transitivity}
\end{equation}

where $c$ and $d$ are representing number of (\#) triangles and triads. When two edges are shared by one common vertex then we call triads. The number of triads present in a network is equivalent to the number of possible triangles. Each triangle have been counted $3$ times (once at each vertex) when we calculated the number of triangles of the whole network. By examining the 3rd column of Table \ref{compare}, we can observe the transitivity values associated with different community networks generated through various methods.

\subsubsection{Average Clustering Coefficient}
In graph theory, the clustering coefficient \cite{watts1998collective} of graph is measure of the degree to nodes in which nodes try to cluster together and the average clustering coefficient \cite{schank2005approximating} is representing the mean of local clustering of any network \cite{kaiser2008mean}. Local clustering can be found for all the vertices $v$ in the graph $\mathcal{G}$ by fractions of triangles that actually exist over all the possible triangles in its neighborhood. Let, node $v$ is containing $T(v)$ triangles and $deg(v)$ is representing the degree of $v$, the mathematically we can define:

\begin{equation}
	c_v = \frac{2T(v)}{deg(v)(deg(v-1))}
\end{equation}

Now, Let $\mathcal{CC}_{avg}$ is representing the average clustering coefficient of graph $G$, then numerically we candefine:

\begin{equation}
	\mathcal{CC}_{avg} = \frac{1}{a} \sum_{v \in G}^{}c_v
\end{equation}

where $a$ is the number of nodes. When we look at the 4th column of Table \ref{compare}, we can see the average clustering coefficient values for different community networks that were created using different methods.

\subsubsection{Degree Distribution}
\begin{figure}[h]
	%\begin{minipage}[t]{.30\textwidth}
		\centering
		\includegraphics[width=\textwidth]{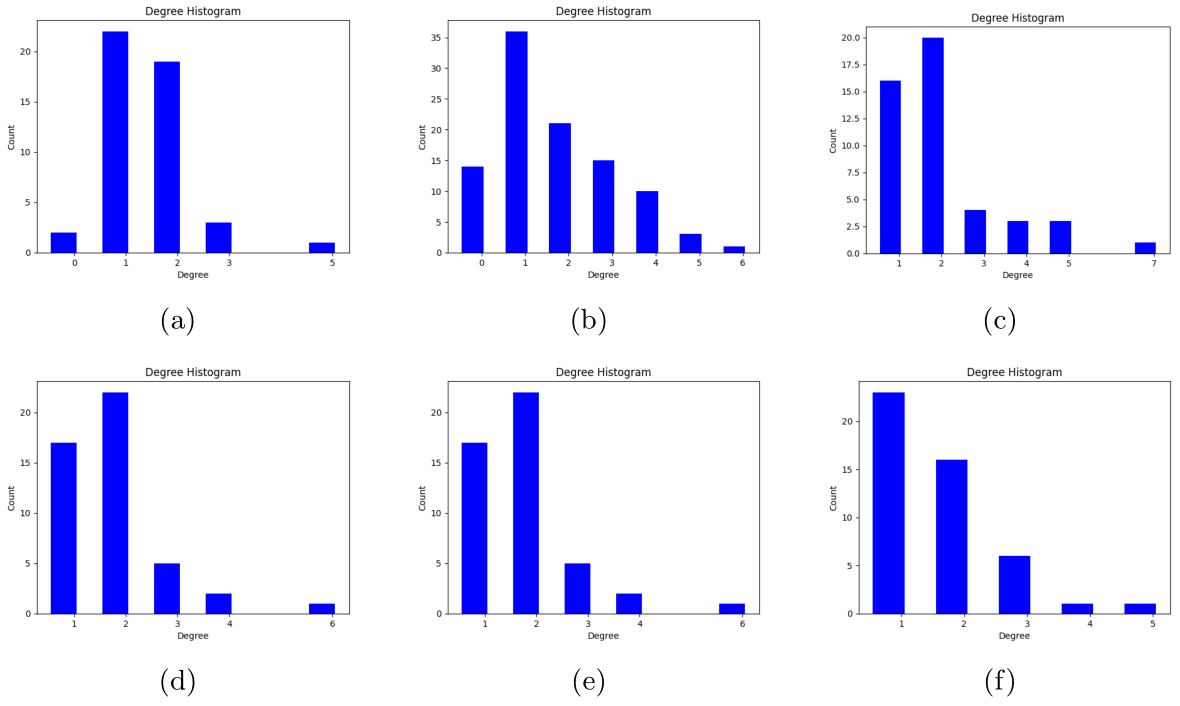}
	%\end{minipage}
		\caption{Degree Histogram: (a) ABCDE (b) ALCDE (c) Grivan-Newman (d) Louvain (e) Leiden (f) Label Propagation}
	\label{degree_histogram}
\end{figure}

After studying Fig. \ref{degree_histogram}, we noticed something interesting about how proteins are connected in different methods within various networks. Most methods showed a similar way of connecting proteins, like having the same network. But there was one method, called AT-ELC, which had a slightly different way of connecting proteins. Understanding these patterns can help us learn more about how these groups function and what makes them special in their networks.

\subsubsection{Subgraph Counts}
Subgraph counting\cite{ribeiro2021survey} is the fundamental approach in several network analysis methodologies, used to compare or categorized the network. We can see the subgraph generated from different methodologies of the Omicron PPI network in Fig. \ref{sub-graph_A-EBC} to Fig. \ref{sub-graph_label}. We can check Table \ref{subgraph_in_all_method} to find the question "Which method is generated how many subgraphs?"

\begin{figure}[h]
	%\begin{minipage}[t]{.30\textwidth}
		\centering
		\includegraphics[width=\textwidth]{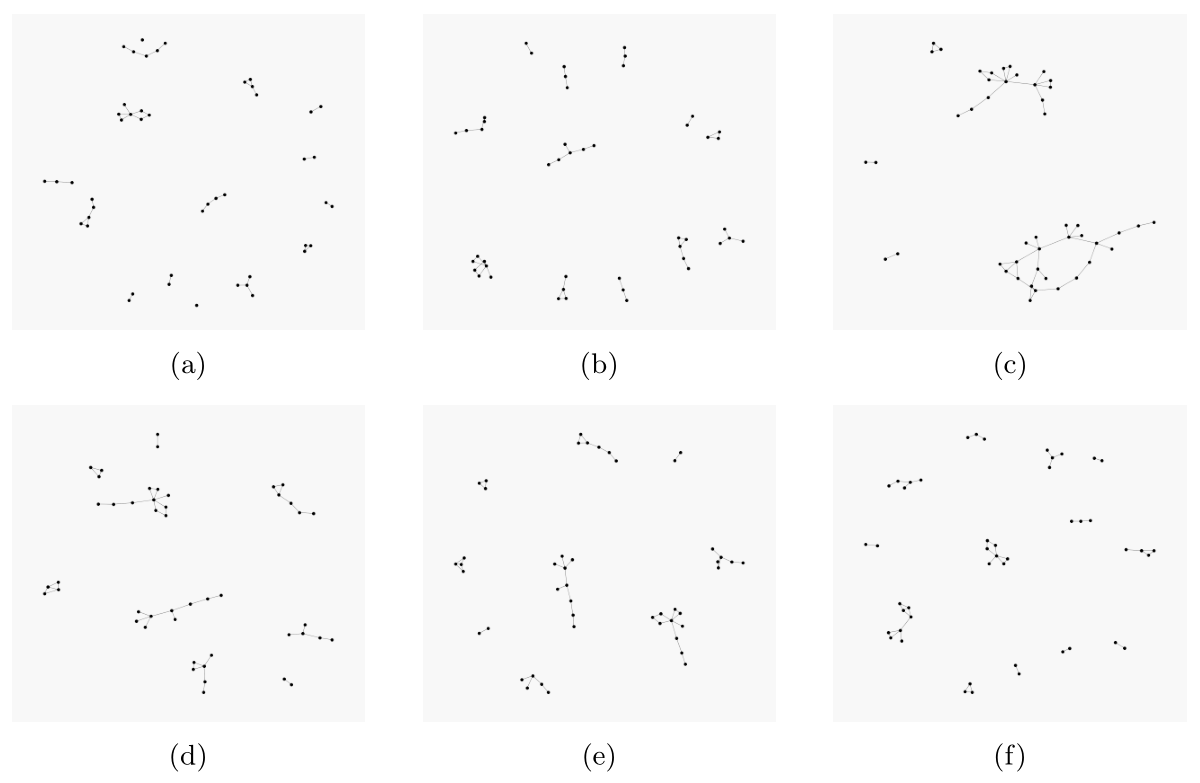}
	%\end{minipage}
	\caption{Comparision of subgraphs in network generated from different methods: (a) ABCDE (b) ALCDE (c) Grivan-Newman (d) Louvain (e) Leiden (f) Label Propagation}
	\label{subgraph}
\end{figure}

\begin{table}[h]
	\caption{No. of subgraph in each methodologis}
	\label{subgraph_in_all_method}
	\centering
	\begin{tabular}{@{}llllll@{}}
		\toprule
		\tiny A-EBC & \tiny A-ELC & \tiny Girvan-Newman &\tiny Louvain & \tiny Leiden & \tiny Label Propogation \\ 
		\midrule
		\tiny 15 & \tiny12 & \tiny5 & \tiny9 & \tiny9 & \tiny13 \\
		\botrule 
	\end{tabular}
\end{table}

\subsubsection{Graphlet Analysis}
\begin{figure}
	\centering
	\includegraphics[width=\textwidth]{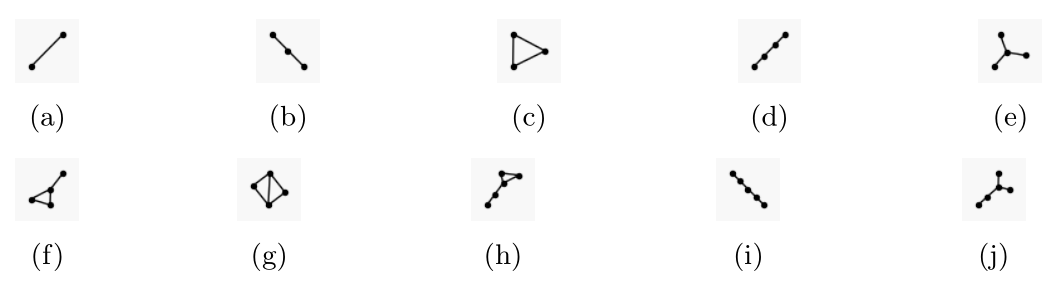}
	\captionof{table}{2 to 5-node graphlets are generated from the omicrone PPI network in different methods: (a) edge (b) 3-path (c) triangle (d) 4-path (e) 4-star (f) 4-tailedtriangle (g) 4-chordalcycle (h) 5-tailedtriangle (i) 5-path (j) 5-tailedstar}
	\label{graphlet}
	%\footnotesize{}
	%Graphlets (2- to 5-node) in undirected, unweighted networks (from ref. 22). The 30 graphlets defined by Pr?ulj et al.18 are labeled G0 to G29. 
\end{figure}

In PPI networks, proteins tend to organize into functional modules, where cellular functions are performed by a small group of interacting proteins. These functional patterns, known as motifs, appear to be preserved across different species. When exploring how a group of nodes collaboratively accomplishes a specific function, we focus on using small graphs, called graphlets, as fundamental units for network comparison, instead of merely matching individual nodes or edges. Graphlet \cite{ahmed2017graphlet}\cite{ahmed2015efficient} and network motifs are widely used metrics to evaluate the structural similarities or distinctions between various community networks. In our research, we have used node graphlets\cite{sarajlic2016graphlet} to find the structural difference between the PPI networks created in different methods. $2$ to $5$ node graphlets are considered to compare the network in our work which are visualized in Tab. \ref{graphlet}. In Tab. \ref{graphlet}, $(a)$ indicates the $2-node$ graphlets, $(b)$ to $(c)$ indicates the $3-node$ graphlets, $(d)$ to $(g)$ indicates the $4-node$ graphlets, and, $(h)$ to $(j)$ indicates the $5-node$ graphlets respectivly (more than 5-node graphlet is count as large graphlet). 

\begin{figure}
	\centering
	\captionof{table}{Overview of graphlet notation}
	\label{graphlets-notation}
	\includegraphics[width=\textwidth]{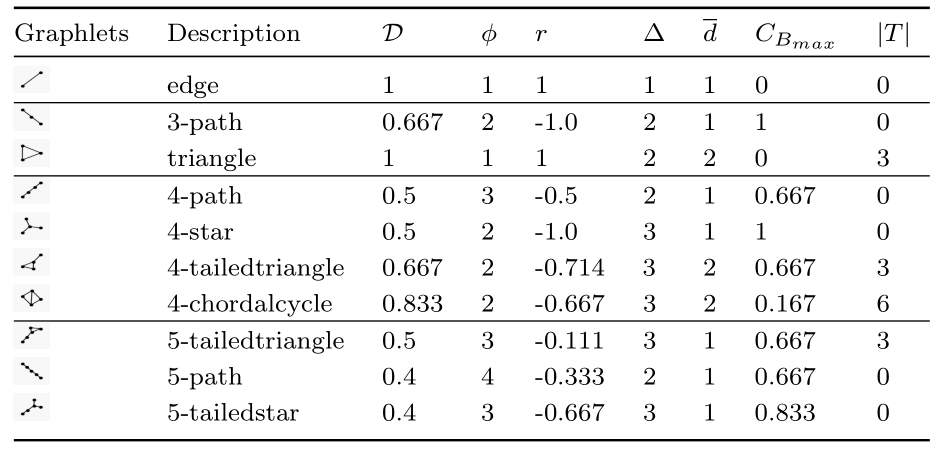}
	%\footnotesize{}
	%Graphlets (2- to 5-node) in undirected, unweighted networks (from ref. 22). The 30 graphlets defined by Pr?ulj et al.18 are labeled G0 to G29. 
\end{figure}

Here, our graphlet size $k=\{2, 3, 4, 5\}$. We have mentioned global properties of mentioned graphlet in Tab. \ref{graphlets-notation}. $\mathcal{D}$ denotes the density, $\phi$ denotes the diameter, $r$ denotes the assortativity, $\Delta$ and $\overline{d}$ denote maximum and minimum degree, $C_{B_{max}}$, and $|T|$ denotes the total triangle of graphlets.

We can see (Fig. \ref{subgraph}) that A-ELC  is containing five edge, one triangle, one $3-path$, one $4-path$, one $4-star$, one $4-tailedtriangle$, and one $5-path graphlets$. The presence of a triangle indicates a tightly interconnected group of three nodes. triangle are essential for measuring clustering and transitivity in the network, showing the potential for information flow and influence propagation within this cohesive group. The occurrence of a $3-path$ suggests a linear subgraph connecting three nodes sequentially. This pattern may represent a sequential relationship or pathway between nodes, providing insight into potential sequential processes or interactions in the network. The presence of a $4-star$ pattern implies a central node with four peripheral nodes. This structure may signify a hub-and-spoke arrangement, where the central node exerts influence over the surrounding nodes. It could indicate a central player or connector in our community network. The occurrence of a $4-tailedtriangle$ indicates a triangle with an additional node forming a tail-like extension. This graphlet highlights the potential presence of hierarchical relationships within the network, where one node influences a group of nodes. The occurrence of a $5-tailedtriangle$ implies a triangle with two additional nodes forming tails. This structure may represent a more complex hierarchical relationship, with two nodes influencing a central triangle. The presence of a $5-path$ indicates a linear subgraph connecting five nodes sequentially. This pattern may suggest a sequential chain of interactions or processes within the network. The A-ELC containing the same type of graphlets only excluding one $4-path$ graphlet. We can conclude that community network of A-EBC and A-ELC both are similar. Grivan Newman method only containing two $edge$ and one $triangle$ in $3-5$ node graphlet. Louvain and Leiden both the method containing two $edge$, one $triangle$, one $4-chordalcycle$ which suggests a cyclic subgraph with four nodes, where an additional edge connects non-adjacent nodes in the cycle. This graphlet indicates the existence of loops or feedback mechanisms in the network, contributing to information flow and stability, and one occurrence of a $5-tailedstar$ (new pattern unlike previous pattern) suggests a star-like subgraph with a central node and five tails. This pattern may indicate the existence of a central node with multiple peripheral nodes connected to it, signifying a potential core-periphery structure in your community network. We can conclude that community network of Louvain and Leiden both are similar.

The Label Propagation incorporates a pattern similar to AT-ELC, excluding the 5-node type graphlet. However, it introduces 5-node type graphlet pattern $5-tailedstar$ like  Louvain and Leiden method. We can see in Fig. \ref{subgraph} to count the different types of graphlets for different methods.

\subsection{Validation of Community Networks}
Modularity is a way to measure how well a network is divided into communities \cite{newman2018networks}\cite{clauset2004finding}. It tells us if the groups of nodes within the network are more interconnected with each other than they would be in a random network. Let an undirected graph $G$ with vertices $V$ and edges $E$, and a community partition $\mathcal{C}$, the modularity ($\mathcal{M}$) is calculated as follows:
\begin{equation}
	\mathcal{M} = \frac{1}{2b} \sum_{i, j}^{}(A_{ij} - \frac{d_i d_j}{2b}) \delta(c_i, c_j)
\end{equation}
where the graph's adjacency matrix is represented by $A$ and the total number of edges is represented by $b$. The entry in the graph's adjacency matrix called $A_{i,j}$ indicates whether there is an edge between nodes $i$ and $j$ and $A_{i,j} = 1$ if $(i, j) \in E$, otherwise $0$. The degrees of nodes $i$ and $j$ are $d_i$ and $d_j$, respectively. The community assignments for nodes $i$ and $j$, respectively, are $c_i$ and $c_j$. 
\begin{equation}
	\delta(c_i, c_j) = 
	\begin{cases}
		1, & \text{if } i, j \in \mathcal{C}\\
		0, & \text{otherwise}
	\end{cases}
\end{equation}
The modularity score ranges from $-1$ to $1$. A positive $\mathcal{M}$ indicates that the network's partition has a higher density of edges within communities than expected by chance. A negative $\mathcal{M}$ suggests that the partition is worse than a random one, and a $\mathcal{M}$ close to $0$ means the community structure is not significantly better or worse than random. So, a higher modularity score indicates a good separation of nodes into distinct communities. In simple terms, modularity assesses the strength of community structure by comparing the actual number of edges within communities to the expected number of edges in a random network. If the actual number of edges within communities is much higher than expected, the modularity score is positive, suggesting a well-defined community structure. On the other hand, if the number of edges within communities is close to what would be expected in a random network, the modularity score is close to zero, indicating a weak community structure. Table \ref{modularity} presents the modularity values for various approaches. %Our proposed approach, A-EBC, achieves the highest modularity score. Leiden and Louvain methods have near to similar modularity values, $0.775$ and $0.717$ respectively. Grivan Newman's modularity is $0.681$, while both A-ECL and Label Propagation methods exhibit the preety good same modularity of $0.664$. 

\begin{table}[h]
	\caption{Modularity in each methodologies}
	\label{modularity}
	\centering
	\begin{tabular}{@{}llllll@{}}
		\toprule
		\tiny AT-EBC & \tiny AT-ELC & \tiny Girvan-Newman &\tiny Louvain & \tiny Leiden & \tiny Label Propogation \\ 
		\midrule
		\tiny 0.885 & \tiny 0.698 & \tiny 0.570 & \tiny 0.847 & \tiny 0.847 & \tiny 0.865 \\
		\botrule 
		%5 & own &  \\ \hline
	\end{tabular}
	
\end{table}
Our proposed approach, AEB-CD, outperforms all other methods in terms of modularity, obtaining the highest score $0.885$ among all the techniques compared. Our other approach AEL-CD scored modularity $0.698$ which is greater than the Grivan Newman method. The Leiden and Louvain methods exhibit similar modularity values $0.847$ indicating comparable performance for these two techniques. Label Propagation methods show higher modularity value than other three existing method. The Grivan Newman method achieves a modularity score of $0.570$, which is lower than the scores obtained by all the other methods. In summary, the modularity values provide insights into the quality of community structures generated by different methods. A higher modularity score indicates a more well-defined and cohesive community partition, while lower scores suggest weaker or less distinct community structures. Based on the results, our proposed approach AEB-CD demonstrates the strongest community detection performance in terms of modularity among the compared methods. The AEL-CD method also exhibits a remarkable capability to generate cohesive and meaningful communities within the network.

%To calculate modularity, we sum up the differences between the actual number of edges within communities and the expected number of edges. The final modularity score takes into account all pairs of nodes in the network and their community assignments.

%Modularity is a valuable metric used in community detection algorithms to evaluate the quality of the communities identified within a network. It helps us understand how well the network is divided into cohesive groups or communities.

\section{Conclusions}
Community detection plays a crucial role in comprehending and assessing the intricate architecture of vast networks. This methodology leverages edge properties in graphs or networks, making it a more fitting choice for network analysis when compared to clustering approaches. Unlike clustering algorithms, which may inadvertently isolate individual peripheral nodes from their rightful communities, community detection algorithms excel in preserving the network's overall structure. A diverse array of algorithms has been proposed and implemented for network community detection. Depending on the features of the particular network and the current problem domain, each of these approaches has particular advantages and disadvantages. The inference of meaningful communities within networks of interacting proteins has emerged as a prominent focus in contemporary biological research. This endeavor holds tremendous potential in unraveling the functional aspects and contextual significance of specific macromolecular assemblies. Furthermore, it can aid in the identification of proteins that potentially influence crucial biological processes. Developing efficient algorithms capable of effectively identifying relevant protein communities within networks is instrumental in advancing drug discovery efforts and enhancing disease treatment, even during early stages. This paper employs a combination of hierarchical, flow simulation-based and semi-supervised clustering methodologies to explore and uncover communities of protein-protein interactions within the Omicron protein-protein interaction network. By utilizing these innovative approaches, the study aims to shed light on the intricate organization of protein interactions and contribute to the understanding of the underlying mechanisms at play.

%\section*{Declarations}

%Some journals require declarations to be submitted in a standardised format. Please check the Instructions for Authors of the journal to which you are submitting to see if you need to complete this section. If yes, your manuscript must contain the following sections under the heading `Declarations':

\begin{itemize}
\item Funding: NA
\item Conflict of interest/Competing interests:
%\item Ethics approval 
%\item Consent to participate
%\item Consent for publication
%\item Availability of data and materials
%\item Code availability: Code will be provide on request.
%\item Authors' contributions
\end{itemize}

{\small
	\bibliographystyle{plain}
	\bibliography{20230718_community}
}

%\bibliography{20230718_community}

\end{document}